\documentstyle[preprint,aps]{revtex}
\tighten
\begin{document}
\draft
\title{A simple functional form for
proton-${}^{208}$Pb
total reaction cross sections}
\author{S.~Majumdar, P.~K.~Deb, and K.~Amos}
\address{School of Physics, University of Melbourne, Parkville,
Victoria, Australia, 3010} 
\date{\today}
\maketitle
\begin{abstract}
A simple functional form has been found 
that gives a good representation
of the total reaction cross sections for the scattering 
from ${}^{208}$Pb of
protons 
with energies in the range 30 to 300 MeV.
\end{abstract}

The values of total reaction cross sections
of nucleon scattering from nuclei (stable 
as well as radioactive) are required
in a number of fields of study.
Some of quite current interest
concern the transmutation of
long-lived radioactive waste into shorter-lived products
using accelerator-driven systems
(ADS) and in predicting dosimetries for patients
in radiation therapy.
To be able to specify those total reaction cross sections
in a simple functional form then has great utility for any associated
evaluation such as of nucleon production in 
spallation calculations.

Recently~\cite{PRLnewplus}
nucleon-nucleus  total reaction cross sections to 300 MeV
for $^{12}$C and $^{208}$Pb have been predicted
in good agreement with data.  Those cross sections were
evaluated using 
$g$-folding optical potentials formed by full folding realistic
two nucleon ($NN$) interactions 
with credible nucleon-based structure models  
of those nuclei.
The effective interactions at each energy in the range
to 300 MeV were defined from the $NN$ $g$ matrices 
(solutions of the Bruckner-Bethe-Goldstone equations)
of the free $NN$ (BonnB) interactions; and so
vary with energy and the medium.
The ground state of ${}^{12}$C
was specified from  a complete $(0+2)\hbar\omega$ shell model
evaluation~\cite{PRLnewplus},
while that of ${}^{208}$Pb was obtained from a Skyrme-Hartree-Fock
calculation~\cite{Brown}.
The resulting $nA$ optical potentials are complex, energy dependent
and very non-local.
All details of the approach
and numerous demonstrations 
of its successful use 
with targets spanning the entire mass range (3 to 238)
are given in the recent review~\cite{Review}.
But each such prediction of nucleon scattering
is an involved calculation that culminates with
use of large scale computer programs,
and those of DWBA98~\cite{Review,Raynal} in our recent
studies.
It would be very utilitarian if aspects of such scattering
were indeed well approximated by a simple convenient
functional form.
We demonstrate herein that for the total reaction
cross sections such a form may exist at least
for heavy nuclei such as ${}^{208}$Pb.

We have used
the $g$-folding optical potential calculations  to specify
proton-${}^{208}$Pb scattering $S$ matrices
at energies $E\propto k^2$,
\begin{equation}
S^{\pm}_l \equiv S^{\pm}_l(k) = e^{2i\delta^{\pm}_l(k)} =
\eta^{\pm}_l(k)e^{2i\Re\left[ \delta^{\pm}_l(k) \right] }\ ,
\end{equation}
where $\delta^\pm_l(k)$ are the (complex) scattering
phase shifts and $\eta^{\pm}_l(k)$ are the moduli
of the $S$ matrices. The superscript designates
$j = l\pm 1/2$. 
Total reaction cross sections then  are predicted by
\begin{eqnarray}
\sigma_R(E)  &=&  \frac{\pi}{k^2} \sum^{\infty}_{l=0} \left\{ (l+1)
\left[ 1 - \left( \eta^+_l \right)^2 \right] + l \left[ 1 - \left(
\eta^-_l \right)^2 \right] \right\}\nonumber\\
&=& \frac{\pi}{k^2}\sum^{\infty}_{l=0} \sigma_l^{(R)}(E)\ .
\label{xxxx}
\end{eqnarray}
The partial reaction cross sections $\sigma_l^{(R)}(E)$ so found 
we now take as 'data'. 
The values from calculations
of the scattering of protons from ${}^{208}$Pb
are displayed in the  Fig.~\ref{Pb_30-300}
by the filled circles for energies 30, 40, 50,
65, 100, 160, 200, 250, and 300 MeV as the peak magnitudes
increase respectively.
They 
suggest to us that the partial total reaction
cross sections $\sigma_l^{(R)}(E)$  
can be described by the simple functional form,
\begin{equation}
\sigma_l^{(R)}(E) = (2l+1) \left[1 + e^{\frac{(l-l_0)}{a}}\right]^{-1}
 + \epsilon (2l_0 + 1)
 e^{\frac{(l-l_0)}{a}}
\left[ 1 + e^{\frac{(l-l_0)}{a}} \right]^{-2}
\label{Fnform}
\end{equation}
with $l_0(E,A)$, $a(E,A)$, and $\epsilon(E,A)$ varying smoothly with 
energy and mass.

The summation giving the total reaction cross section
can be limited to a value $l_{max}$ and the associated
form tends appropriately to the 
known high energy limit.  With increasing energy,
$l_{max}$ becomes so large that  the
exponential fall-off of the functional form, Eq.~\ref{Fnform},
can be approximated as a straight vertical line
($l_0 = l_{max}$).
In that case, the total reaction cross section
equates to the area of a triangle, and
\begin{equation}
\sigma_R \Rightarrow \frac{\pi}{2 k^2} l_{max}(2l_{max} + 1)  
 \approx \frac{\pi}{k^2} l_{max}^2\ .
\end{equation}
Then with $l_{max} \sim kR$, at high energies
\begin{equation}
\sigma_R \Rightarrow \pi R^2\ ;
\end{equation}
the geometric cross section as required.

The fits of this functional form to the "data" from ${}^{208}$Pb
are displayed by the solid curves in Fig.~\ref{Pb_30-300} with
the values of the three parameters given at the specific energies
connected by straight lines in 
Fig.~\ref{Pbparams}.  These values are plotted as functions
of $\sqrt{E}$ (effectively $k$) and it is clear that
they are smooth variations with energy.  Of these the linearity
of $l_0$ with $\sqrt{E}$ is most evident and we find
\begin{equation}
l_0(E, {}^{208}{\rm Pb}) 
\sim kR - 3\ ;\ \ \ R \sim 1.6 A^{-\frac{1}{3}}\ .
\end{equation}
The best fit values of these parameters are listed in 
Table~\ref{paramtab}.


For any dynamical situation in which numerous partial 
reaction cross sections contribute non-negligibly to the sum
in Eq.~\ref{xxxx}, one can use the limit form
\begin{equation}
\sigma_R(E) 
= \frac{\pi}{k^2}\sum^{\infty}_{l=0} \sigma_l^{(R)}(E)
\Longrightarrow \frac{\pi}{k^2} \int_0^\infty \sigma^{(R)}(E; \lambda)
\ d\lambda\ .
\end{equation}
Then with $x = \frac{\lambda - l_0}{a}$
and using the basic integrals 
\begin{eqnarray}
\int_{-l_0/a}^\infty \frac{1}{1 + e^x} dx
&=& 
\left[ \int_0^{l_0/a} \frac{1}{1 + e^{-x}} dx
+ \int_0^\infty \frac{e^{-x}}{1 + e^{-x}} dx\right]
= \frac{l_0}{a} + \sum_{n=1}^\infty (-)^{n+1} e^{-\frac{l_0}{a}n}
\nonumber\\
\int_{-l_0/a}^\infty \frac{x}{1 + e^x} dx
&=& \left[ -\int_0^{l_0/a} \frac{x}{1 + e^{-x}} dx
+ \int_0^\infty \frac{xe^{-x}}{1 + e^{-x}} dx\right]
\nonumber\\
&&\hspace*{2.0cm}=
-\frac{1}{2} \frac{l_0^2}{a^2}
+ 1.645 + \sum_{n=1}^\infty
(-)^n \left( 1 + \frac{1}{n^2}\right) e^{-\frac{l_0}{a}}
\ ,
\end{eqnarray}
the integral observable is 
\begin{eqnarray}
\sigma^{(R)}(E) &=& \frac{\pi}{k^2} \sum_l \sigma_l^{(R)}(E)
\to \frac{\pi}{k^2} \int_0^\infty \sigma^{(R)}(\lambda;E) \
d\lambda
\nonumber\\
&=& \frac{\pi}{k^2} \left\{
l_0(l_0+1) + 3.29 a^2 + a\epsilon (2l_0+1)
\right.\nonumber\\
&&\hspace*{2.5cm}\left.
+ \sum_{n=0}^\infty e^{-n\frac{l_0}{a}}
(-)^n \left[
2a^2 \left\{ 1+ \frac{1}{n^2}\right\} +
(2l_0+1)\left\{ a\epsilon - \frac{1}{n} \right\}
\right] \right\}
\nonumber\\
&&{\mathop \to_{n=1}}\ 
\frac{\pi}{k^2}
\left\{
l_0(l_0+1) + 3.29 a^2 + a\epsilon (2l_0+1)
+ \left[(2l_0 + 1) \left\{\frac{1}{n} - a\epsilon\right\}
- 4a^2  \right] e^{-\frac{l_0}{a}}
\right\}
\label{sig_fn_tot}
\end{eqnarray}
Thus if  $l_0$, $a$, and $\epsilon$  are well behaved functions from 
matching to observed data, one can have confidence that
these simple prescriptions give reliable estimates 
of reaction cross section values that have not been measured
as yet but may be required as input in, for example,
spallation production calculations.
The ratios of the total reaction cross sections
calculated under this
approximation to those
determined from the microscopic optical model
potentials
are listed in the last column of the Table.
The results compare well to within a few percent.

\begin{table}
\caption[]{Parameter values giving fits to $\sigma_l^{(R)}(E)$}
\begin{tabular}{|c|ccc|c|}
\hspace*{1.5cm}E (MeV)\hspace*{1.5cm} &
\hspace*{1.0cm} $l_0$\hspace*{1.0cm} &
\hspace*{1.0cm} $a$\hspace*{1.0cm} &
\hspace*{1.0cm} $\epsilon$\hspace*{1.0cm}&
$\sigma^{(R)}_{n=1}/\sigma^{(R)}(E)$ \\
\hline
\ 30 & 10.2&1.123& -1.19 & 0.99\\
\ 40 &11.9&1.02& -1.13 & 1.02\\
\ 50 &13.1&0.89& -0.81 & 1.01\\
\ 65 &14.6&0.96& -0.63 & 1.00\\
100 &18.6&1.59&  -0.92 & 0.99\\
160 &24.4&2.65&  -1.25 & 0.97\\
200 &27.5&3.06&  -1.35 & 0.96\\
250 &30.8&3.41&  -1.48 & 0.98\\
300 &34.0&3.75&  -1.67 & 0.97\\
\end{tabular}
\label{paramtab}
\end{table}


\begin{figure}
\caption[]{The total partial reaction cross sections for proton scattering from
${}^{208}$Pb.}
\label{Pb_30-300}
\end{figure}

\begin{figure}
\caption[]{The parameters of the simple functional form
for the p-${}^{208}$Pb partial total cross sections.}
\label{Pbparams}
\end{figure}




\begin{references}

\bibitem{PRLnewplus}
P. K. Deb, K. Amos, S. Karataglidis, M. B. Chadwick,
 and D. G. Madland,
Phys. Rev. Lett. {\bf 86}, issue 15 (2001).




\bibitem{Brown} B. Alex Brown,
Phys. Rev. Lett.  {\bf 85}, 5296 (2000).

\bibitem{Review}
K.~Amos, P.~J. Dortmans,
H.~V. von Geramb, S.~Karataglidis, and
J.~Raynal, Adv. in Nucl. Phys.
{\bf 25}, 275 (2000).


\bibitem{Raynal}
J. Raynal, 
{\emph{computer code DWBA98}}
NEA 1209/05, 1999.


\end{references}
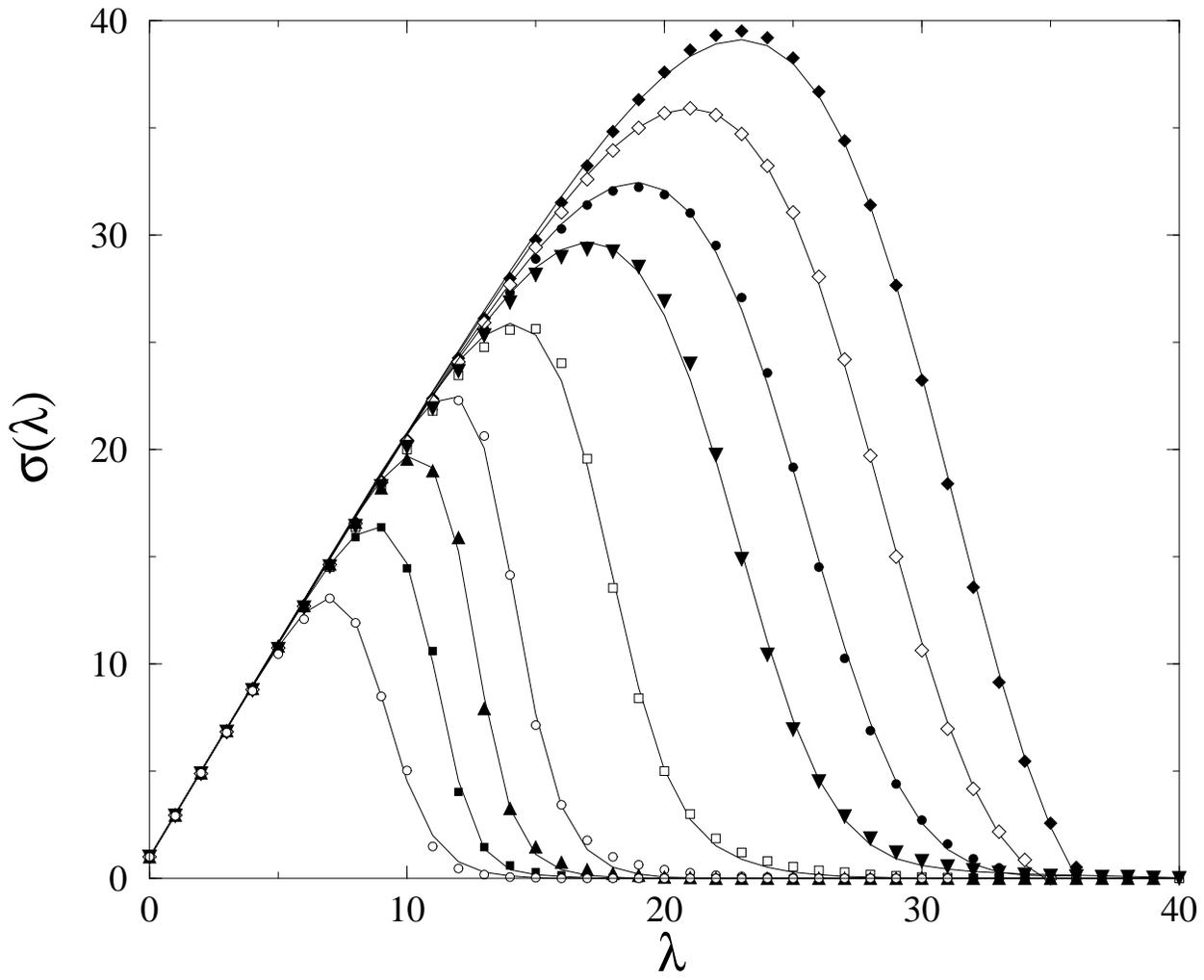
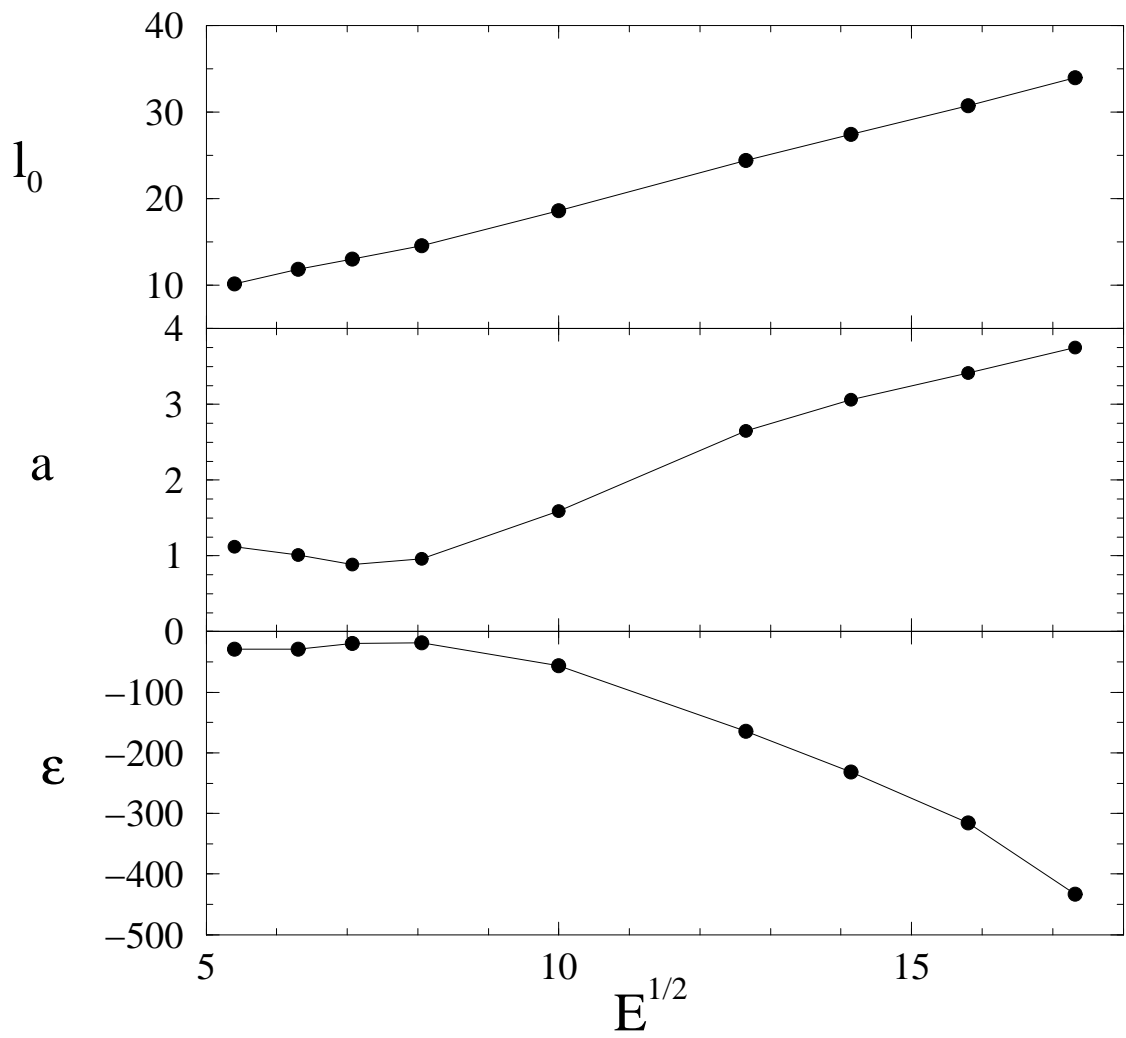
\end{document}